\documentstyle[osa,manuscript]{revtex}
\begin{document}
\title{Remanent magnetization of high-temperature Josephson junction arrays}
\author{W. A. C. Passos, P. N. Lisboa-Filho and W. A. Ortiz}
\address{Grupo de Supercondutividade e Magnetismo, Departamento de F\'{i}sica\\
Universidade Federal de S\~{a}o Carlos, Caixa Postal 676 - 13565-905 S\~{a}o%
\\
Carlos, SP, Brazil}
\maketitle

\begin{abstract}
% DON'T CHANGE THIS LINE
In this work we study the remanent magnetization exhibited by tridimensional
disordered high-T$_{c}$ Josephson junction arrays excited by an AC magnetic
field. The effect, as predicted by numerical simulations and previously
verified for a low-T$_{c}$ array of Nb, occurs in a limited range of
temperatures. We also show that the magnetized state can be excited and
detected by two alternative experimental routines.
\end{abstract}

% INITIALIZE - DONT CHANGE
%
%
%

% \author{}   % Use this and the next line only if there is a second
% \address{Another University, etc.}  % address. (Remove the left % marks)
%

\bigskip 

\bigskip In a recent work\cite{passos} we have demonstrated that Josephson
junction arrays (JJAs) fabricated from granular Nb may exhibit a magnetic
remanence, $M_{r}$, upon excitation by a magnetic field. As predicted\cite
{araujo-moreira}, the magnetized state occurs in a window of temperatures,
whose extent depends on the critical current, $J_{c}$, of the junctions.
Also, there is a threshold value for the magnetic field in order to drive
the JJA to the state where flux is retained after suppression of the field
\cite{passos}$^{,}$\cite{araujo-moreira}. In Ref. 1 we have also shown that
the profile of $M_{r}$ is sensitive to the critical current dispersion, what
stresses the prospective use if this effect as a suitable tool for
determining the critical current distribution of the array, $N(J_{c})$.

This contribution presents selected parts of a systematic study of the
remanent magnetization displayed by our newly produced high-temperature
tridimensional disordered arrays (3D-DJJAs), fabricated from granular YBa$%
_{2}$Cu$_{3}$O$_{7-\delta }$. The experimental results confirm the
predictions, revealing that the remanence develops in a limited range of
temperature.

Granular YBCO material used to fabricate the arrays was prepared employing a
modified method of polymeric precursors\cite{kakihana}. This route consists
of mixing oxides and carbonates in stoichiometric amounts dissolved in HNO$%
_{3}$, and then to an aqueous citric acid solution. A metallic citrate
solution is then formed, to which ethylene glycol is added, resulting in a
blue solution which was neutralized to pH\symbol{126}7 with ethylenediamine.
This solution was turned into a gel and subsequently decomposed to a solid
by heating at 400 $^{o}$C. The sample was heat-treated at 850 $^{o}$C for 12
h in air with several intermediary grindings, in order to prevent
undesirable phase formations. Then, it was pressed into a pellet using
controlled uniaxial (5,000 kgf/cm$^{2}$) pressure and sintered at 950 $^{o}$%
C for 6 h in O$_{2}$. This pellet is a 3D-DJJA, in which the junctions are
weakly coupled grains, i.e., weak-links (WLs) formed by a sandwich of YBCO
grains and intergrain material. As a consequence of the uniaxial pressure,
samples produced in this way are anisotropic, a feature that can be either
enhanced, by using higher pressures, or reduced, by applying isostatic
pressures. Also, thermal treatment plays a fundamental role on creation and
control of WLs and anisotropy, as will be thoroughly discussed elsewhere\cite
{passos2}.

The sample studied here exhibits all characteristic features of a genuine
3D-JJA, the most significant of which are shown in Fig.1: the main picture
is a low-field measurement of a positive magnetization\cite{moment}
(Wohlleben Effect), for $H=0.02\;Oe$. The inset displays a Fraunhofer
pattern for the real part of the magnetic susceptibility $\chi _{AC}$. As
demonstrated in Ref. 6, this is an indirect determination of $J_{c}$.

To study the remanent state of the arrays, we employ two routines especially
developed for detection and study of granular JJAs: the Temperature Scan
Routine (TS) and the Field Scan Routine (FS). The core of both experimental
procedures consists of two steps:

i. the sample is submitted to an AC field ($h$) consisting of a train of
sinusoidal pulses, after what h is kept null;

ii. with $h=0$, the magnetic moment of the sample is measured.

In the FS routine we measure $\chi _{AC}(h)$ performing steps (i) and (ii)
as h is varied at a fixed temperature. On the other hand, the TS routine is
employed to measure, at a fixed value of $h$, $\chi _{AC}(T)$ through steps
(i) and (ii). All measurements were performed using a Quantum Design MPMS-5T
SQUID magnetometer. Both routines were extensively explored, furnishing
valuable results for the purposes of this work. In this short paper,
however, we emphasize the similarities among results obtained employing the
two alternative routines. Remaining parts of this study, including many
other aspects of the problem, will be published elsewhere.

As expected, the high-T$_{c}$ disordered array studied here, a heat-treated
isotropic 3D-DJJA of YBCO, exhibits the predicted magnetic behavior. The
magnetized state at zero field can be easily recognized from measurements
using either one of the above mentioned routines. Remanence versus
temperature curves normalized to peak values, are shown in Fig.2. The main
graph represents a direct measurement of $M_{r}(T)$ on warming, employing
the TS routine. Intragranular contributions were not subtracted from this
curve, but are totally irrelevant, as we have certified by measuring
unlinked grained material, for which no magnetic remanence was detected. As
in the case of the Nb array reported previously\cite{passos}, the remanence
is intense for a limited temperature interval. In the present case, the
temperature window for which the array is magnetically active has a quite
long low-T tail, revealing that $N(J_{c})$ is broad. It should be noticed
that the magnetic response ceases at $T^{\ast }=83\;K$, a temperature
significantly smaller than $T_{c}=90\;K$. The inset shows a few
representative points of an experiment performed using the FS routine for $%
h_{o}=10\;mOe$. The AC field was cycled up to $3.8\;Oe$ and then down to
zero, after which values of $\Delta M_{r}$ were calculated by subtraction
between $M_{r}(h_{o})$ increasing and decreasing $h$. The curve obtained is
a replica of that measured directly using the TS routine.

In conclusion, we have measured the predicted magnetic remanence of JJAs,
using a 3D-DJJA fabricated from granular YBCO. The remanence is intense
within a limited interval of temperatures. The $M_{r}(T)$ profile, which is
sensitive to the critical current dispersion, reveals a fairly broad $%
N(J_{c})$ for the array.

%
% ({\it REVTEX} 3.0 automatically issues
% a \newpage command when the \begin{table} or \begin{figure}
% commands are used, so the figures and tables will be placed
% on separate pages by {\it REVTEX}).

\begin{figure}[tbp]
\caption{Positive magnetization (Wohlleben Effect) of a 3D-JJA of YBCO,
measured on cooling at $H=30mOe$. Inset shows the Fraunhofer pattern of $%
\protect\chi _{AC}$ for the sample.}
\end{figure}

\begin{figure}[tbp]
\caption{Magnetic remanence exhibited by a 3D-DJJA of YBCO, as measured
using the TS (main graph) and the FS (inset) routines.}
\end{figure}


\begin{references}
\bibitem{passos}  W. A. C. Passos, F. M. Araujo-Moreira and W. A. Ortiz, to
be published in J. Appl. Phys. (Cond.Mat. 9910118).

\bibitem{araujo-moreira}  F. M. Araujo-Moreira et al., Phys. Rev. Lett. 78
(1997) 4625.

\bibitem{kakihana}  M. Kakihana, Journal of Sol-Gel Technology 6, (1996) 7.

\bibitem{passos2}  W. A. C. Passos et al., ''Experimental realization of
tridimensional Josephson junction arrays'', to be submitted.

\bibitem{moment}  Throughout the paper, M is the volume magnetization, i.e.,
the magnetic moment of the samples.

\bibitem{barbara}  P. Barbara et al., Phys. Rev. B 60 (1999) 7489.
\end{references}
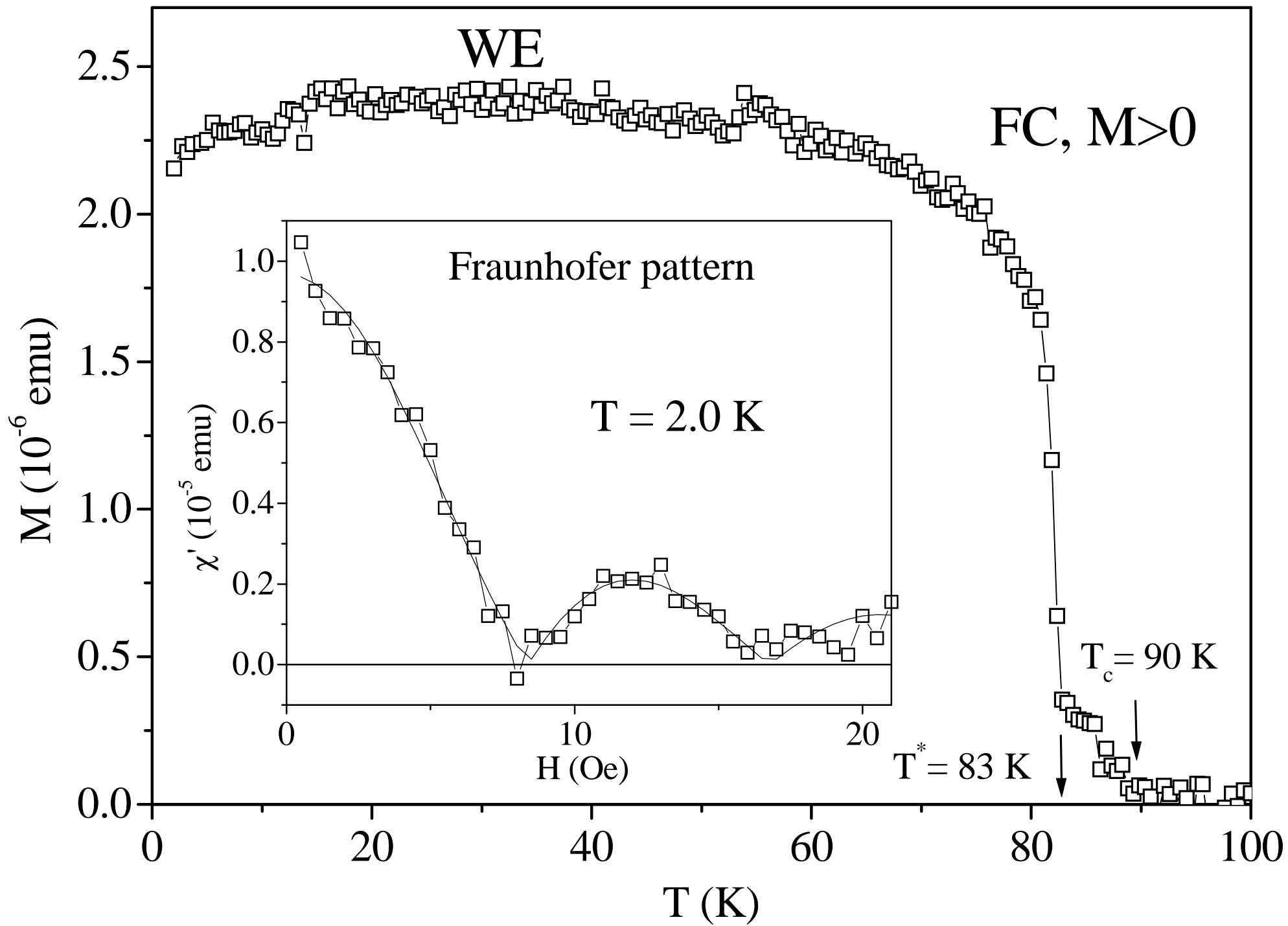
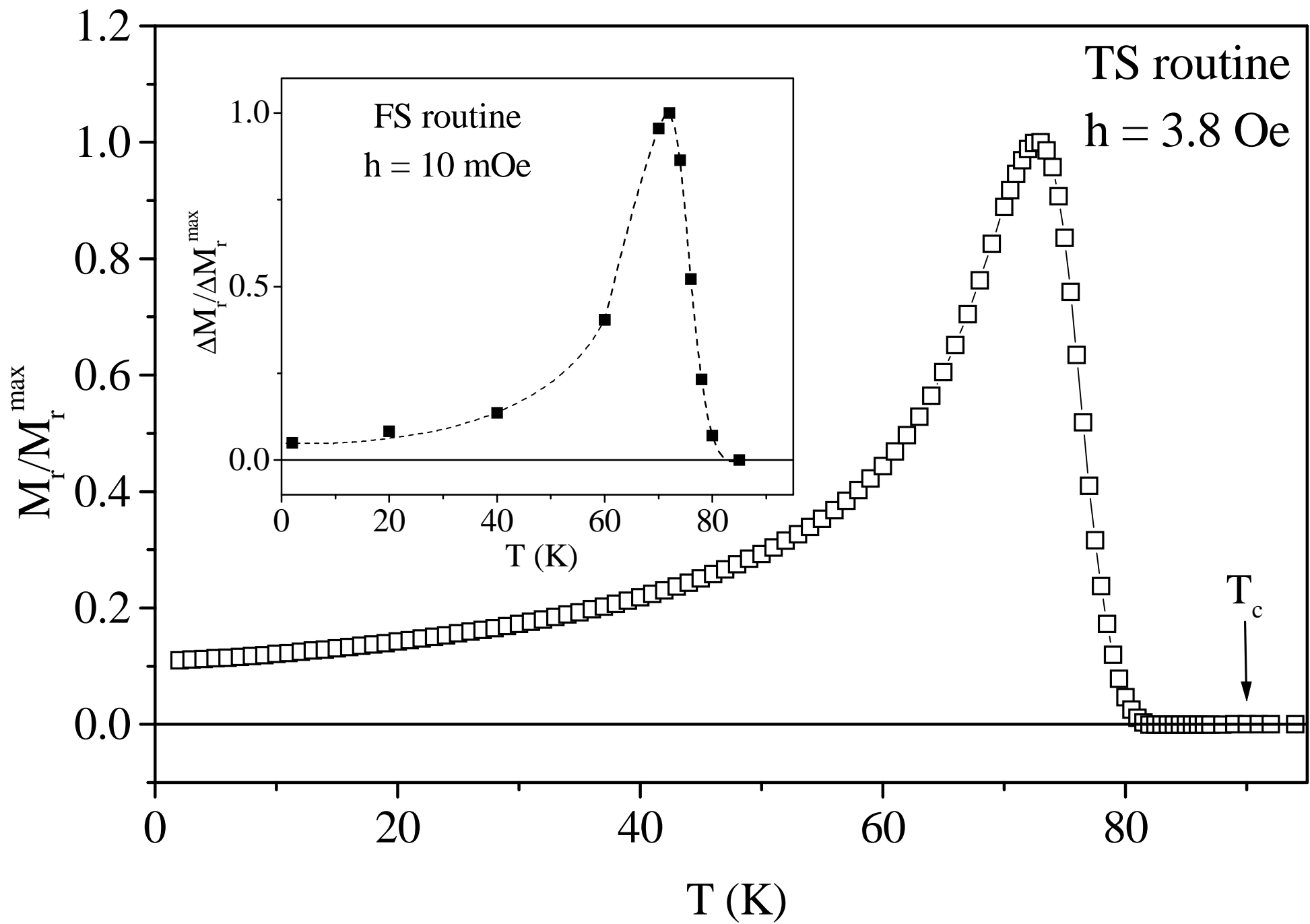
\end{document}